\documentclass[aps,prl,showpacs,floatfix,twocolumn,superscriptaddress,longbibliography]{revtex4-1}

\usepackage[pdftex]{graphicx}
\usepackage{color}
\usepackage{hyperref}
\usepackage{verbatim}
\usepackage{soul}
\usepackage{glossaries}
\usepackage{upgreek}
\usepackage{todonotes}

\def\mathbi#1{\ensuremath{\textbf{\em #1}}}
\def\Q{\mathbi{Q}}
\def\QCO{\ensuremath{\mathbi{Q}_{\mathrm{CO}}}}
\def\QSO{\ensuremath{\mathbi{Q}_{\mathrm{SO}}}}
\def\TCO{\ensuremath{T_{\mathrm{CO}}}}
\def\TSO{\ensuremath{T_{\mathrm{SO}}}}
\def\TC{\ensuremath{T_{\mathrm{cycle}}}}
\def\um{\ensuremath{\upmu\text{m}}}

\newacronym{RIXS}{RIXS}{Resonant inelastic X-ray scattering}
\newacronym{XAS}{XAS}{X-ray absorption spectrum}
\newacronym{XPCS}{XPCS}{X-ray photon correlation spectroscopy}
\newacronym{LSNO}{LSNO}{La$_{2-x}$Sr$_{x}$NiO$_{4+\delta}$}
\newacronym{LSCO}{LSCO}{La$_{2-x}$Sr$_{x}$CuO$_{4}$}
\newacronym{LBCO}{LBCO}{La$_{2-x}$Ba$_{x}$CuO$_{4}$}
\newacronym{CO}{CO}{charge order}
\newacronym{SO}{SO}{spin order}
\newacronym{HWHM}{HWHM}{half-width at half-maximum}
\newacronym{CSX}{CSX}{Coherent Soft X-Ray}
\newacronym{HTT}{HTT}{high temperature tetragonal}
\newacronym{LTO}{LTO}{low temperature orthorhombic}
\newacronym{EPC}{EPC}{electron-phonon coupling}

\begin{document}

\title{Charge Condensation and Lattice Coupling Drives Stripe Formation in Nickelates}

\author{Y. Shen}\email[]{yshen@bnl.gov}
\affiliation{Condensed Matter Physics and Materials Science Department, Brookhaven National Laboratory, Upton, New York 11973, USA}

\author{G. Fabbris}
\affiliation{Condensed Matter Physics and Materials Science Department, Brookhaven National Laboratory, Upton, New York 11973, USA}
\affiliation{Advanced Photon Source, Argonne National Laboratory, Lemont, Illinois 60439, USA}

\author{H. Miao}
\affiliation{Condensed Matter Physics and Materials Science Department, Brookhaven National Laboratory, Upton, New York 11973, USA}
\affiliation{Material Science and Technology Division, Oak Ridge National Laboratory, Oak Ridge, Tennessee 37830, USA}

\author{Y. Cao}
\affiliation{Condensed Matter Physics and Materials Science Department, Brookhaven National Laboratory, Upton, New York 11973, USA}
\affiliation{Materials Science Division, Argonne National Laboratory, Lemont, Illinois 60439, USA}

\author{D. Meyers}
\affiliation{Condensed Matter Physics and Materials Science Department, Brookhaven National Laboratory, Upton, New York 11973, USA}
\affiliation{Department of Physics, Oklahoma State University, Stillwater, Oklahoma 74078, USA}

\author{D. G. Mazzone}
\affiliation{Condensed Matter Physics and Materials Science Department, Brookhaven National Laboratory, Upton, New York 11973, USA}
\affiliation{Laboratory for Neutron Scattering and Imaging, Paul Scherrer Institut, CH-5232 Villigen, Switzerland}

\author{T. Assefa}
\affiliation{Condensed Matter Physics and Materials Science Department, Brookhaven National Laboratory, Upton, New York 11973, USA}

\author{X. M. Chen}
\affiliation{Condensed Matter Physics and Materials Science Department, Brookhaven National Laboratory, Upton, New York 11973, USA}

\author{K. Kisslinger}
\affiliation{Center for Functional Nanomaterials, Brookhaven National Laboratory, Upton, New York 11973, USA}

\author{D. Prabhakaran}
\author{A. T. Boothroyd}
\affiliation{Department of Physics, University of Oxford, Clarendon Laboratory, Oxford, OX1 3PU, United Kingdom}

\author{J. M. Tranquada}
\affiliation{Condensed Matter Physics and Materials Science Department, Brookhaven National Laboratory, Upton, New York 11973, USA}

\author{W. Hu}
\author{A. M. Barbour}
\author{S. B. Wilkins}
\author{C. Mazzoli}
\affiliation{National Synchrotron Light Source II, Brookhaven National Laboratory, Upton, New York 11973, USA}

\author{I. K. Robinson}
\author{M. P. M. Dean}\email[]{mdean@bnl.gov}
\affiliation{Condensed Matter Physics and Materials Science Department, Brookhaven National Laboratory, Upton, New York 11973, USA}

\date{\today}

\begin{abstract}
Revealing the predominant driving force behind symmetry breaking in correlated materials is sometimes a formidable task due to the intertwined nature of different degrees of freedom. This is the case for La$_{2-x}$Sr$_{x}$NiO$_{4+\delta}$ in which coupled incommensurate charge and spin stripes form at low temperatures. Here, we use resonant X-ray photon correlation spectroscopy to study the temporal stability and domain memory of the charge and spin stripes in La$_{2-x}$Sr$_{x}$NiO$_{4+\delta}$. Although spin stripes are more spatially correlated, charge stripes maintain a better temporal stability against temperature change. More intriguingly, charge order shows robust domain memory with thermal cycling up to 250~K, far above the ordering temperature. These results demonstrate the pinning of charge stripes to the lattice and that charge condensation is the predominant factor in the formation of stripe orders in nickelates.
\end{abstract}

\maketitle

Emergent phenomena in strongly correlated materials arise due to multifarious interactions among charge, spin and lattice degrees of freedom. Such complexity hampers the ability to understand their remarkable states and realize new functionalities \cite{Chumak2015spintronics}. Identifying dominant interaction is, however, challenging as different interactions act simultaneously and can yield complex ground states with more than one form of order \cite{Fradkin2015}. A representative phenomenon of this type is the electronic stripes that appear in various strongly correlated materials \cite{Mori1998LMO, Lee2006review, Ulbrich2012review, Comin2016review}. These effects have been considered extensively in cuprate high-temperature superconductors, which host charge and sometimes spin stripe order, typically with a simple factor-of-two relationship between the charge and spin incommensurabilities  \cite{Tranquada1995cuprate, Mook2000YBCO, Hoffman2002Bi2212}. Nickelates also host both superconductivity and stripe order \cite{Tranquada1994LNO, Tranquada1996LSNO, Li2019Nisuperconductor}, but no system has yet been shown to simultaneously host both orders. The existence of stripe order in La$_4$Ni$_3$O$_8$, which appears rather similar to superconducting Nd$_{1-x}$Sr$_x$NiO$_2$ \cite{Zhang2016LaNiO438, Zhang2017Ni438, Zhang2019LaNiO438, Lin2020strong}, does, however, support the likely proximity of stripe order and superconductivity. While static stripe order appears to suppress bulk 3D superconductivity, some researchers have suggested that stripe fluctuations may act to promote superconductivity \cite{Emery1997spin, Kivelson1998electronic, Agterberg2020PDW_review}. Therefore, understanding the driving forces behind charge and spin stripe formation and dynamics in strongly correlated materials has attracted considerable attention and may be crucial to understanding unconventional superconductivity. Stripe formation has been studied in the past through detailed measurements of stripe transition temperatures and correlation lengths \cite{Chen1993LSNO, Cheong1994LSNO, Lee1997LSNO, Yoshizawa2000LSNO, Lee2001LSNO, Kajimoto2001LSNO, Ghazi2004LSNO, Freeman2004LSNO, Raczkowski2006LSNO} and associated Landau model analysis \cite{Wochner1998LSNO, Zachar1998Landau}. The problem has also been addressed via model Hamiltonian analysis that suggested that lattice coupling might be crucial to stabilize stripes \cite{Zaanen1994freezing, Hotta2004LSNO}. The implementation of resonant \gls{XPCS} at modern low-emittance synchrotron sources opens new routes to directly probe stripe formation and dynamics \cite{Chen2016LBCO, Thampy2017LBCO, Chen2019LBCO, Ricci2019LSNO}. 

Herein, we report the first resonant \gls{XPCS} experiment to simultaneously probe \gls{CO}, \gls{SO} and lattice coupling in a stripe-ordered material, focusing on the prototypical material \gls{LSNO} with $x=0.225$ $\delta=0.07$. Although \gls{SO} is more correlated and stable at 70~K, \gls{CO} is more robust in temporal stability against temperature changes, which we attribute to \gls{EPC}. This is further supported by our discovery that the \gls{CO} domains are effectively pinned to the lattice and the corresponding speckle patterns remain highly reproducible with thermal cycling up to 250~K, well above the transition temperature {\TCO}. \gls{SO}, however, is not directly coupled to the lattice and loses its domain memory once the sample is warmed across the magnetic transition temperature {\TSO}. These results imply that charge condensation, and its coupling to the lattice and disorder, is the driving force behind stripe ordering.

\begin{figure}[t]
\includegraphics{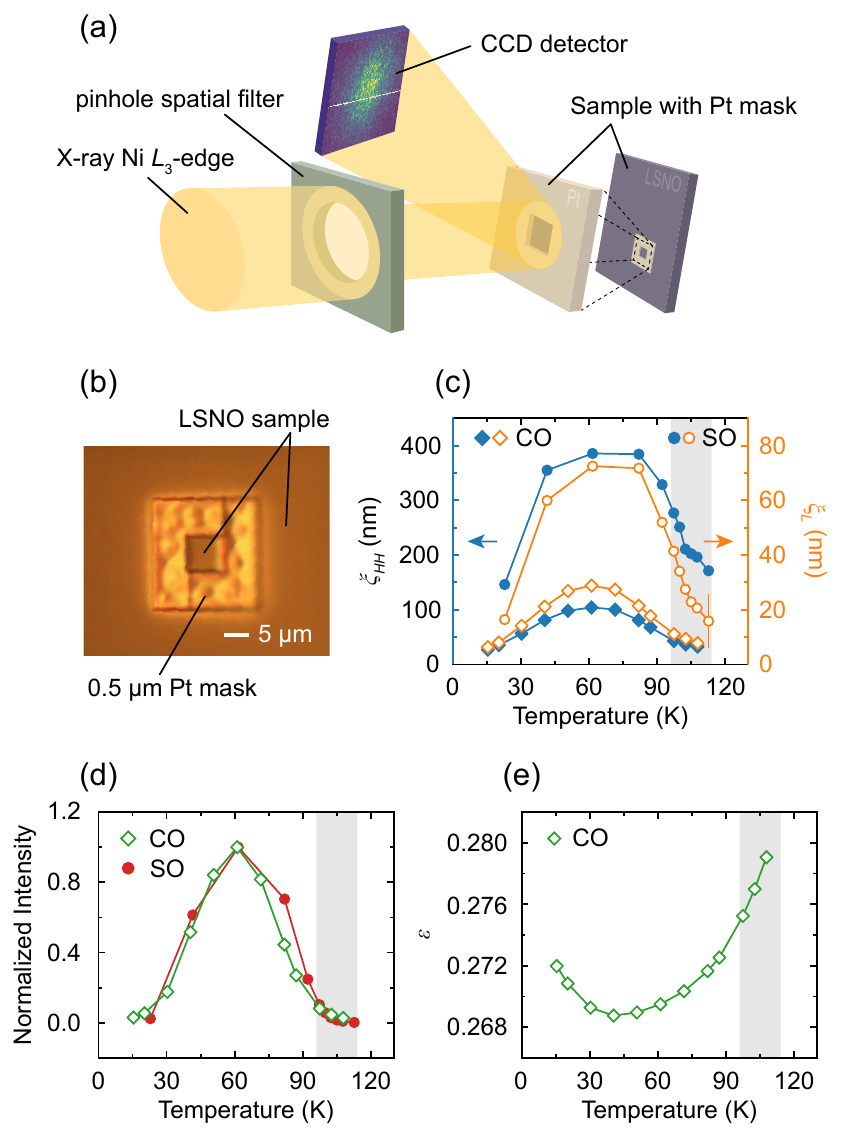}
\caption{Experimental configuration and \acrfull{CO} and \acrfull{SO} superlattice peaks. (a) The instrumental setup for the measurements at CSX. The X-ray beam is set to the Ni $L_3$-edge energy and tuned in order to maximize the strength of the \gls{CO} and \gls{SO} intensity \cite{supp}. It then propagates through the pinhole and is scattered by the \acrfull{LSNO} sample onto the detector. For the domain memory study, a 0.5 {\um} thick Pt mask was deposited on the sample \cite{supp}. (b) An optical micrograph of the Pt mask on the (110) surface of \gls{LSNO} single crystal. (c) Temperature dependence of the correlation lengths along [$H$, $H$, 0] and [0, 0, $L$] directions. The correlation length is defined as $\xi = d/\textrm{HWHM}$ where $\textrm{HWHM}$ stands for half-width at half-maximum in reciprocal lattice units and $d$ is the unit cell size in the appropriate direction \cite{incommensurability}. (d) Temperature dependence of the peak heights evaluated from fitting of the \gls{CO} and \gls{SO} superlattice peaks, which are normalized according to their values at 60~K. The signals were fitted with a three-dimensional Lorentzian function. (e) Incommensurability defined by the peak position of \gls{CO} {\Q} vector as a function of temperature. The shaded areas indicate the onset temperature range for \gls{CO} and \gls{SO}.}
\end{figure}

X-ray measurements were carried out at the \gls{CSX} 23-ID-1 beamline at the National Syncrotron Light Source II with X-ray energy tuned to the Ni $L_3$-edge (Fig.~1a). The \gls{LSNO} single crystal was synthesized by the floating-zone method with a Sr concentration of $x = 0.225$ \cite{Prabhakaran2002growth}. As shown later, the \gls{CO} incommensurability is $\epsilon \approx 0.27$, larger than $x$, which is likely related to oxygen doping since $\delta = 0.07$ \cite{Freeman2006magnetization}. The sample's surface normal was close to the [$H$, $H$, 0] direction. Thus, we made ($H$, $H$, $L$) the scattering plane and focused on peaks with ${\QCO} = (\epsilon, \epsilon, 1)$ and ${\QSO} = (1/2-\epsilon/2, 1/2-\epsilon/2, 0)$ \cite{supp}. The reciprocal lattice units (r.l.u.) is defined in terms of {\Q} = ($H$, $K$, $L$) = $(2\pi/a, 2\pi/b, 2\pi/c)$ within the space group I4/mmm and $a = b = 3.84$~{\AA}, $c = 12.65$~{\AA}. For the domain memory measurements, we used a 0.5~{\um} thick Pt mask, which had been deposited on the sample in order to reproducibly illuminate the same sample volume independent of possible thermal drifts in the sample position (Fig.~1b) \cite{supp}.

We start by characterizing the superlattice peaks corresponding to \gls{CO} and \gls{SO} at different temperatures using standard resonant X-ray diffraction. With decreasing temperature, the peak heights first increase substantially through the transition temperatures along with enhanced correlation lengths for both \gls{CO} and \gls{SO} (Fig.~1c, d). Below $\sim$70~K, the peak heights drop and the spatial correlations are relaxed, consistent with previous reports \cite{Hatton2002LSNO, Ghazi2004LSNO, Schlappa2009LSNO}. The reason for this is not uniquely determined, but it may be connected to a spin reorientation at lower temperature \cite{Freeman2004LSNO} or the influence of spin exchange interactions \cite{Ghazi2004LSNO}. Throughout the temperature range, the correlation lengths along [$H$, $H$, 0] direction are much larger than those along [0, 0, $L$] and \gls{SO} possesses a larger correlation length than \gls{CO} (Fig.~1c). Due to the critical fluctuations and short-range correlations near the phase transitions, the onset temperatures, {\TCO} and {\TSO}, are not uniquely defined. We estimate them both to occur between 96 and 114~K. Regarding the incommensurability, the inter-site Coulomb repulsion tends to stabilize $\epsilon$ equal to the hole concentration \cite{Sachan1995LSNO}, while the commensurability effect optimizes stripe formation at $x = 1/3$. The actual incommensurability is a compromise of these two factors \cite{Yoshizawa2000LSNO}. With increasing temperature, thermal fluctuations are expected to start to outcompete Coulomb repulsion \cite{Hatton2002LSNO, Ishizaka2004LSNO, Miao2019LBCO}, driving the incommensurability closer to 1/3 at higher temperature (Fig.~1e).

\begin{figure}[t]
\includegraphics{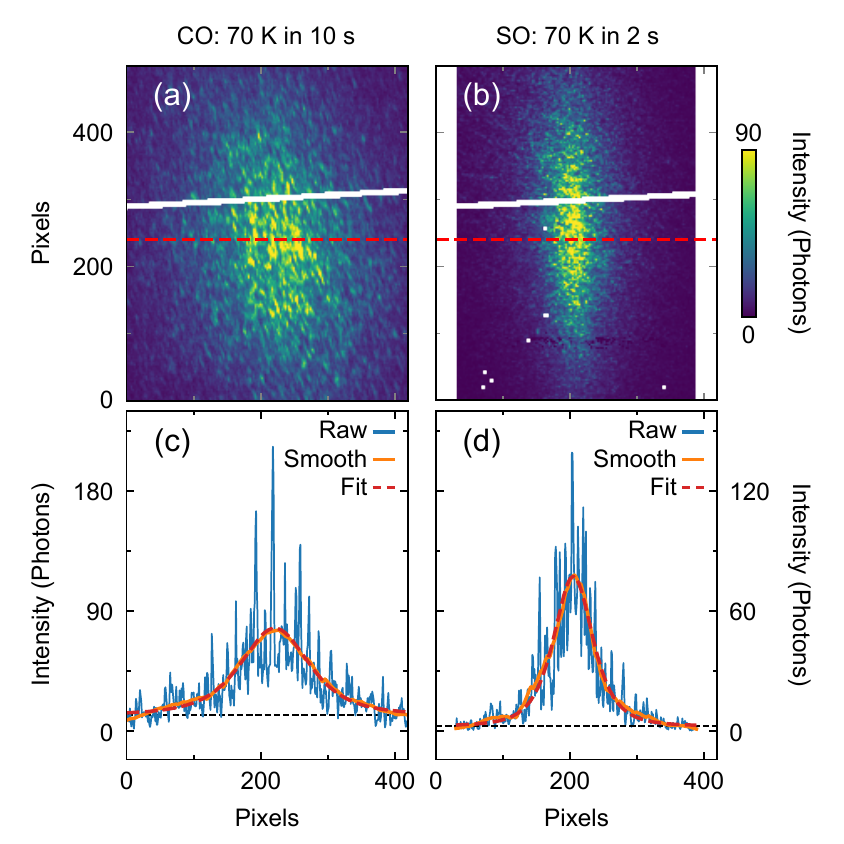}
\caption{Speckle patterns of \gls{CO} and \gls{SO}. (a),(b) Representative detector images around the \gls{CO} and \gls{SO} superlattice peaks measured with a 10~{\um} pinhole. The white pixels arise from the beamstop or detector errors and are omitted from the data. (c),(d) Line cuts through the horizontal red dashed lines in (a) and (b). The envelope of the peak is estimated by smoothing and fitting processes that are shown as red and orange lines, respectively. The black dashed lines are uniform fluorescent background evaluated from fittings.}
\end{figure}

To elucidate the temporal stabilities of \gls{CO} and \gls{SO}, we employ \gls{XPCS} to study the domain distribution and its fluctuations. In \gls{XPCS}, the coherent photons scattered by different domains interfere with each other, leading to a complex ``speckle'' pattern modulated by the usual diffraction lineshape \cite{Brauer1995XPCS, Shpyrko2014XPCS, Chen2016LBCO, Thampy2017LBCO, Ricci2019LSNO, Lee2021Dimensionality}. Figure 2a, b shows the representative speckles of the \gls{CO} and \gls{SO} superlattice peaks at 70~K. The shape of the peak envelope is determined by the spatial correlations and instrument geometry. In particular, the horizontal width of the \gls{SO} peak is mainly determined by the correlations along the [-1, 1, 0] direction while the vertical width is dominated by $c$ axis correlations, elongating the envelope vertically. For the \gls{CO} peak, the vertical width has less contribution from $c$ axis correlations so that the envelope appears more isotropic. Meanwhile, the distribution of the underlying stripe domains is encoded in the positions of the speckles \cite{Chen2019LBCO}, and the shape of the speckles is determined by the Fourier transform of the beam footprint projected onto the detector. The non-zero $L$ component of the \gls{CO} peak makes the footprint of the beam more anisotropic. To show the speckle modulation more clearly, we present in Fig.~2c, d the line cuts through the red dashed lines in Fig.~2a, b. The peak envelope is estimated by two independent methods: smoothing with the Savitzky-Golay filter and fitting with a squared Lorentzian function. The sharp speckle modulation observed here indicates that the fluctuations for \gls{CO} and \gls{SO} are slower than the time windows of the measurements, which is 1~s at 70~K \cite{supp}. Otherwise the contrast of the interference patterns will be significantly reduced \cite{Chen2016LBCO}.

\begin{figure}[t]
\includegraphics{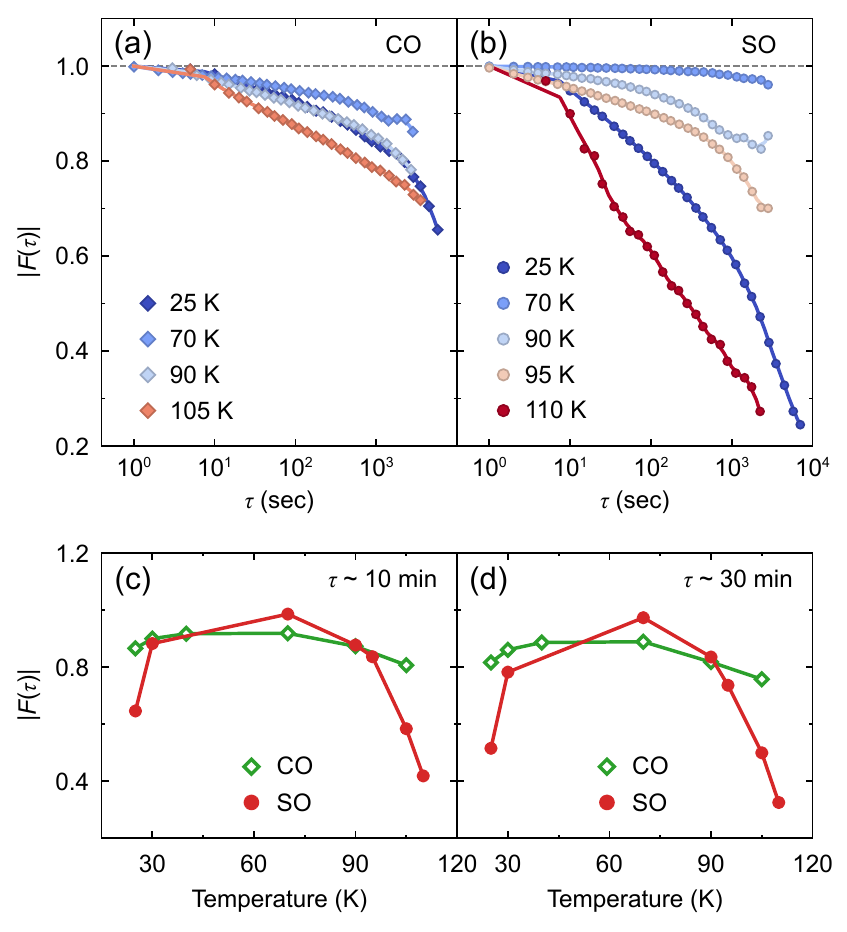}
\caption{Temporal stability of \gls{CO} and \gls{SO}. (a),(b) Time dependence of the intermediate scattering functions at different temperatures. The solid lines are guides to the eye. (c),(d) The scattering functions after certain time delays.}
\end{figure}

\begin{figure*}[t]
\includegraphics{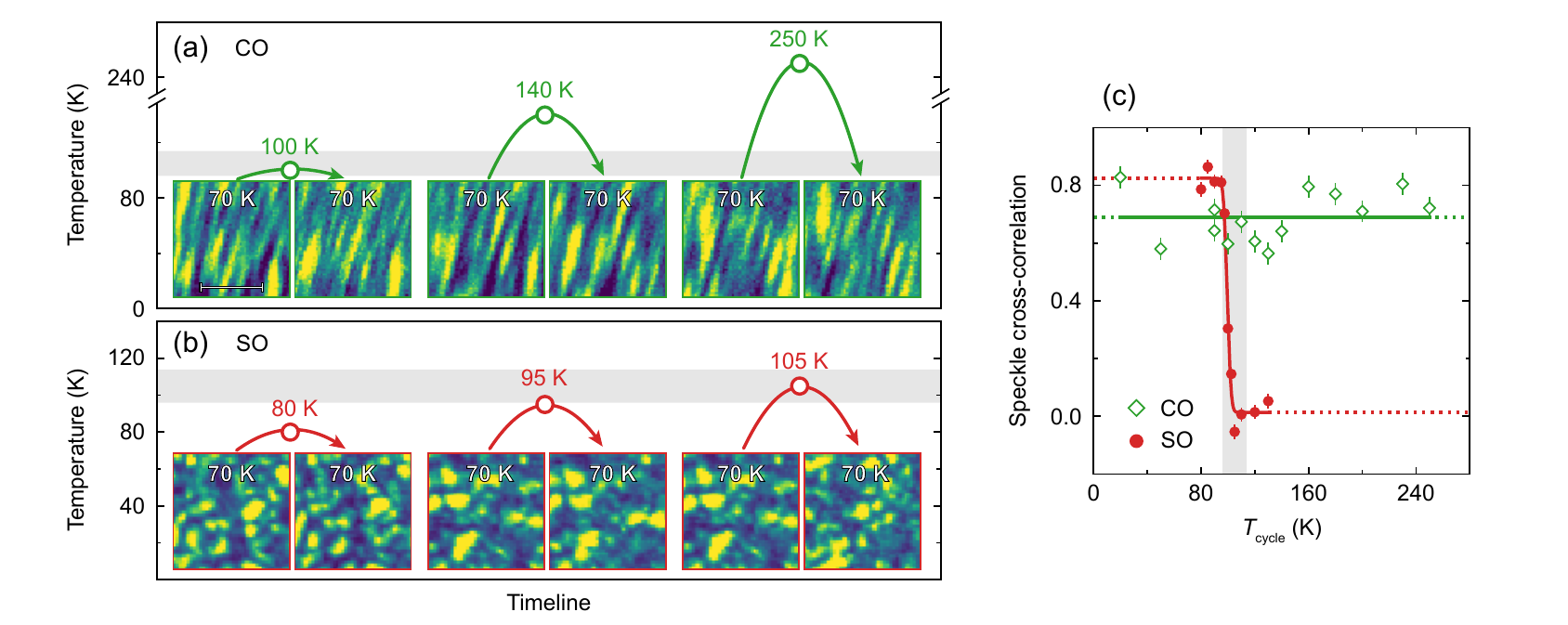}
\caption{Domain memory in \gls{CO} but not \gls{SO}. (a),(b) Representative speckle images before and after thermal cyclings which are indicated by the curved arrows. The open circles stand for the cycling temperatures, {\TC}. For each measurement, we collected images at 70~K, changed the temperature to {\TC} and waited for 10 minutes. Then the sample was cooled back to 70~K and equilibrated for 30 minutes before collecting another image. For both the heating and cooling processes, the temperature ramping rate was fixed to 4~K/min. The white bar in the first speckle image indicates $10^{-3}$ \AA{}$^{-1}$. (c) Temperature dependence of the normalized speckle cross-correlation function, $\xi_{\mathrm{CC}}$. The solid and dashed lines are guides to the eye. The shaded area indicates the range of \gls{CO} and \gls{SO} transition temperatures.}
\end{figure*}

In order to quantify the fluctuation timescale, we measure the time dependence of the speckle patterns and calculate the normalized one-time correlation function \cite{Chen2016LBCO}
\begin{align}
g_2 (\tau) = \frac{\langle I(t)I(t+\tau)\rangle}{\langle I(t)\rangle^2} = 1 + \beta|F(\tau)|^2,
\end{align}
where $I$ represents the total intensity including background, $\tau$ is the lag time and $\langle\ldots\rangle$ stands for the time and ensemble average. The time-dependent evolution can be extracted from the intermediate scattering function, $|F(\tau)|$, which describes the correlation of the speckle patterns separated by a certain time delay. In a statically ordered system, $|F(\tau)|$ will remain unchanged while speckle dynamics causes it to drop as a function of time delay. Distinct from \gls{LBCO}, in which the \gls{CO} is static over a timescale of at least two hours \cite{Chen2016LBCO, Thampy2017LBCO}, $|F(\tau)|$ in \gls{LSNO} decays after several minutes for both \gls{CO} and \gls{SO}, indicating charge and spin dynamics (Fig.~3). Moreover, we find that \gls{CO} and \gls{SO} are both most stable around 70~K when they have longest correlation lengths, but \gls{SO} is more stable than \gls{CO} at 70~K. Although stripes involve a co-modulation of both charge and spin \cite{Zachar1998Landau}, we observe that these have different thermal evolution. As temperature is driven away from 70~K, the temporal stability for \gls{SO} decreases faster, indicating that \gls{SO} is less stable against temperature changes. A qualitatively, but not quantitatively similar trend in \gls{SO} was reported recently in Ref.~\cite{Ricci2019LSNO}. The longer timescales observed here may reflect sample discrimination in strontium and oxygen compositions or improved coherent flux and stability at \gls{CSX} compared to the Advanced Light Source.

From simple energetic considerations, if an order is less temporally stable and has shorter correlation lengths one would expect it to be more fragile to thermal disturbance. The unexpected robustness of \gls{CO} against temperature changes indicates that \gls{CO} is coupled to other degrees of freedom which constrain the \gls{CO} domains during/after the charge condensation (Fig.~3). Such hypotheses can be examined more deeply in term of domain pinning memory effects. Since the speckle positions are primarily determined by the positions of the ordering domains, the comparison of speckle patterns collected at 70~K before and after cycling the sample temperature to {\TC} can evaluate whether the domain distributions are reproduced \cite{Chen2019LBCO}. The usage of Pt mask further ensures that the illuminated sample volume is fixed throughout the thermal cycling (Fig.~1b). It turns out that the speckle patterns of \gls{CO} are rather similar with {\TC} up to 250~K, well above {\TCO} (Fig.~4a). The \gls{SO} speckles, however, change their positions once {\TC} crosses {\TSO} ($\sim 100$~K) (Fig.~4b). This effect can be quantified by calculating the normalized cross-correlation function $\xi_{\mathrm{CC}}$ which describes the similarity between two speckle patterns \cite{Chen2019LBCO, supp}. $\xi_{\mathrm{CC}}$ approaches zero when the two speckle images are different while two identical images will give $\xi_{\mathrm{CC}}$ of one. Correspondingly, we calculate $\xi_{\mathrm{CC}}$ for both \gls{CO} and \gls{SO} speckle patterns with different {\TC} (Fig.~4c). The results again show that \gls{CO} domain distributions are essentially unchanged after thermal cycling to a temperature far above {\TCO} while \gls{SO} speckle pattern loses reproducibility after the system is driven into the disordered state.


The domain memory effect of \gls{CO} is caused by coupling to the host lattice. Local potentials arising from structural disorder induced, for example, by Sr doping, structural domain boundaries or octahedral tilts, provide nucleation centers for the \gls{CO} domains and effectively pin the domains during stripe condensation. Since the average lattice structure of \gls{LSNO} has translational symmetry over a lengthscale smaller than \gls{CO} wavelength it cannot, in self, pin the \gls{CO} domains into reproducible locations. In charge-ordered cuprate \gls{LBCO}, the speckle pattern of \gls{CO} domains loses memory after the sample is heated across the transition temperature from the \gls{LTO} phase into the \gls{HTT} phase \cite{Chen2019LBCO}. Thus, it is expected that the pinning landscape for \gls{CO} in \gls{LBCO} is constrained by twin boundaries created by the \gls{LTO} structural distortion. In \gls{LSNO}, the lattice remains in the \gls{HTT} phase and no long-range \gls{LTO} distortion is observed \cite{Hucker2004PhaseDiagram}. However, short-range stripe-related distortions have been reported to persist up to high temperatures \cite{Simon2013LSNO}. It is possible that either these distortions, or local defects due to Sr-related doping disorder, determine the pinning landscape of \gls{LSNO} in a similar manner.


The pinning effect of \gls{CO} to the structural disorder also evinces the relevance of \gls{EPC} in nickelates, which has been illustrated by the discovery of phonon anomalies and nematic behaviors in \gls{LSNO} \cite{Pashkevich2000stripe, Tranquada2002LSNO, Merritt2020phonon, Zhong2017LSNO}. It has been argued theoretically that without \gls{EPC} \gls{CO} will remain dynamic and not order \cite{Hotta2004LSNO}. For structure-driven \gls*{CO}, phonons soften to zero energy and drag the valence charge along with it to form spatial modulations. Here, however, phonons are softened by a maximum of 20\% \cite{Tranquada2002LSNO}, and charge stripes are formed to reduce Coulomb interactions. \gls*{EPC} helps pin pre-formed charge stripes according to the lattice symmetry, promoting the static \gls{CO}. The presence of \gls{EPC} further couples the \gls{CO} domains to structural disorder, which strengthens the \gls{CO} against thermal fluctuations. Consequently, when \gls{CO} and \gls{SO} lose correlations progressively upon heating or cooling away from 70~K, the fluctuations of \gls{CO} speckles increase more slowly (Fig.~3).

\gls{SO} behaves in a different way. During the formation of \gls{SO}, the spins can align either parallel or antiparallel to their quantization axis. This would disrupt the reproducibility of \gls{SO} speckles after thermal cycling across {\TSO} even if the domain walls are in the same place (Fig.~4). Moreover, the rotational degree of freedom provides an additional fluctuation channel to the ordered spins, facilitating the loss of \gls{SO} stability when driven away from 70~K (Fig.~3). This is in line with the observation of spin reorientation in \gls{LSNO} at low temperatures \cite{Lee2001LSNO, Freeman2004LSNO, Freeman2006magnetization}.


The robustness of \gls{CO} stability and its pinning to the lattice demonstrate that the stripe order in \gls{LSNO} is charge driven. This directly verifies prior theoretical predictions based on Landau theory of coupled charge and spin order parameters \cite{Zachar1998Landau} and may reflect that stripe-order is charge-driven in general. Our approach will be extendable to other materials and even to other degrees of freedom such as orbital order, bringing a powerful means to disentangle the formation mechanisms of intertwined ground states.

\begin{acknowledgements}
This material is based upon work supported by the U.S. Department of Energy (DOE), Office of Basic Energy Sciences. Work at Brookhaven National Laboratory was supported by the U.S. DOE, Office of Science, Office of Basic Energy Sciences, under Contract No.~DE-SC0012704. The work at Argonne National Laboratory was supported by the U.S. Department of Energy, Office of Basic Energy Sciences, under Contract No. DE-AC0206CH11357.  D.~G.~M.\ acknowledges funding from the Swiss National Science Foundation, Fellowship No.\ P2EZP2\_175092. This research used resources at the. 23-ID-1 beamline of the National Synchrotron Light Source, a U.S. DOE Office of Science User Facility operated for the DOE Office of Science by Brookhaven National Laboratory under Contract No. DE-AC02-98CH10886.
\end{acknowledgements}

\bibliography{refs}

\begin{thebibliography}{58}
\bibitem [{inc()}]{incommensurability}%
  \BibitemOpen
  \href@noop {} {}\bibinfo {note} {$\xi = a/\sqrt{2}\Delta H$ where $\Delta H$
  is half-width at half-maximum \ change in $\Delta H$ for a $[H, H, 0]$
  scan}\BibitemShut {NoStop}%
\end{thebibliography}
\end{document}


\title{Supplemental Material: Charge Condensation and Lattice Coupling Drives Stripe Formation in Nickelates}

\renewcommand{\thepage}{S\arabic{page}} 
\renewcommand{\thesection}{S\arabic{section}}  
\renewcommand{\thetable}{S\arabic{table}}  
\renewcommand{\thefigure}{S\arabic{figure}}

\maketitle

This document provides additional details of the X-ray measurements, sample preparation and the wavevectors of \gls{CO} and \gls{SO} peaks.

\section{X-ray measurements}

All X-ray measurements were carried out at the \gls{CSX} 23-ID-1 beamline at the National Synchrotron  Light Source II (NSLS-II) that is dedicated to resonant coherent soft x-ray studies. NSLS-II was used in its standard mode with 1320 electron buckets and 2~ps between the buckets \cite{nsls2url}. \gls{CSX} at NSLS-II has rather ideal properties for these experiments as it delivers an extremely high coherent flux of $\sim10^{13}$ photons/s at the sample. This is, at the time of writing, state-of-the-art in time-averaged coherent flux even including current x-ray free electron lasers, which, despite their very high peak flux, have similar or lower time-averaged flux.

\begin{figure}
\includegraphics{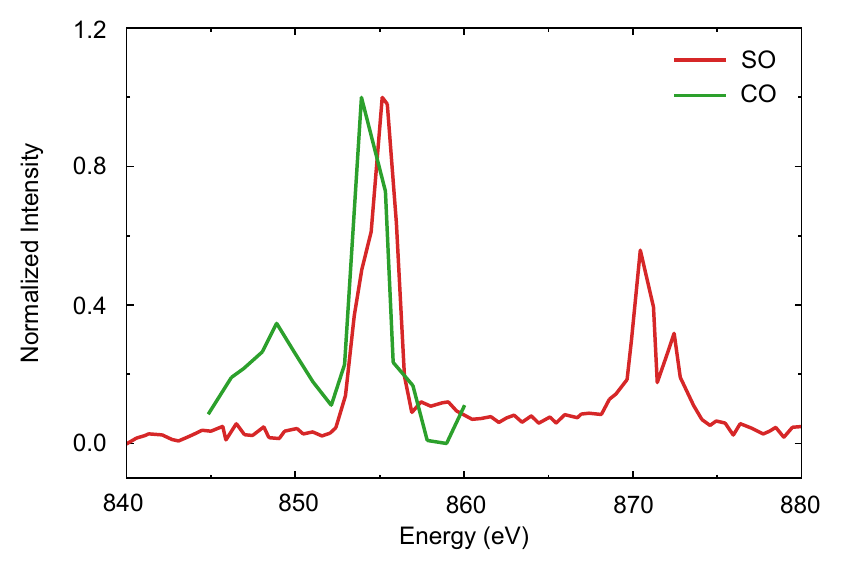}
\caption{Intensities of the superlattice peaks of \gls{CO} and \gls{SO} as a function of phonon energy.}
\end{figure}

To enhance the signals from \gls{CO} and \gls{SO}, we tuned the incident energy to find the Ni $L_3$-edge resonance for both orders separately, as these occur at slightly different energies \cite{Schussler2005LSNO}. Figure S1 shows the resonant profile of the \gls{CO} and \gls{SO}. These results are consistent with prior resonant x-ray scattering on La$_{2-x}$Sr$_x$NiO$_4$ \cite{Schussler2005LSNO}. An overview of resonant soft x-ray scattering is available in Ref.~\cite{Fink2013resonant}.  The X-ray polarization was also chosen, using the \gls{CSX} elliptically polarized undulator, to maximize the signal. As has been proven previously, \gls{SO} is optimized via $\pi$ X-ray polarization and \gls{CO} is similarly strong with either $\sigma$ or $\pi$ polarization \cite{Schussler2005LSNO}. Data were collected using a fast CCD with $30 \times 30$~\um$^2$ pixel size that is 340~mm from the sample. This gives a reciprocal space resolution of $3.8\times10^{-5}$~\AA$^{-1}$/pixel.

In standard \gls{XPCS}, the time resolution is determined by how frequently the detector is read out. This is chosen considering (a) the speed of the dynamics (b) read-out noise from the detector, (c) the minimal time required to achieve sufficient signal, and (d) the maximum read-out rate of the detector (in our case every 10~ms). As described in the main text, we can exclude any significant fluctuations faster than about 10~s, since fast dynamics would reduce the speckle contrast via averaging over the measured time frame \cite{Chen2016LBCO}. We chose the CCD read out time for the measurements based mainly on considerations (a)-(c). The CCD was read out every 1~s at 70~K. In this regime we detect of order 100 photons per pixel and 1 million photons/frame and cover well over $>90$\% of the total \gls*{CO} and \gls*{SO} Bragg peak intensity. We therefore have high confidence that the signals represent the domain dynamics on length scales down to $\sim100$~nm, which is the characteristic lengthscale of the smallest \gls*{CO} and \gls*{SO} domains we measure. At temperatures with weaker scattering the read-out time was increased to up to 5~s.

After any temperature change, the experiment, and in particular the cryostat, was left to equilibrate for a minimum of 30 minutes following previously established procedures to stabilize the setup \cite{Chen2016LBCO, Chen2019LBCO, Kukreja2018orbital}.

\section{sample preparation}

To realize a clean surface for the measurement, we polished the sample in water with sequence of P250 (sandpaper), P600 (sandpaper), 1 micron and 0.25 microns diamond. All steps were performed for about 1~min except for the last one which lasted 2 to 3~min.

In the measurement of domain memory, the sample position could drift during the sample cooling and warming process. To make sure that the same sample volume was illuminated in spite of the possible drift, we deposited a mask on the sample. This was done using a Field Electron and Ion (FEI) Helios 600 dual-beam focused-ion-beam (FIB) at the Center for Functional Nanomaterials (CFN). We deposited four 12~{\um} by 7~{\um} Pt pads with a thickness of 500~nm to form a square pinhole of approximately 5~{\um} by 5~{\um} in size inside a mask of about 19~{\um} by 19~{\um}. The x-ray beam was passed through a 10~{\um} conventional pinhole approximately 5~mm upstream of the sample to form a beam of about 11~{\um} in diameter at the sample, overfilling the pinhole, but not the mask. As the Pt deposition was performed in high-vacuum, possible oxidation to the surface of the sample was avoided. Since the penetration depth of Pt at the Ni $L_3$-resonance is 72~nm, the Pt layer blocks x-ray very efficiently. After the temperature cycle, the sample position was scanned to re-align the beam to the center of the pinhole. The mask was not used during the measurements of the temporal stability of the speckle patterns since the beamline and sample were sufficiently stable in these circumstances and we benefited from using the full intensity of the beam. 

\section{Charge and spin order wavevectors}

The stripes in \gls{LSNO} run diagonally to the Ni-O bonds. We prepared the single crystal with a surface normal close to the [$H$, $H$, 0] direction in order to access a ($H$, $H$, $L$) scattering plane. This plane contains the \gls{CO} and \gls{SO} peaks at wavevectors of {\QCO} = $(N\pm\epsilon, N\pm\epsilon, L)$ with $L$ being odd and {\QSO} = $(N+1/2\pm\epsilon/2, N+1/2\pm\epsilon/2, L)$ with $L$ being even or odd, where $N$ is an integer, $\epsilon$ is the incommensurability parameter. In this way, {\QSO} occurs at detector angle of $2\theta=155^{\circ}$ and an approximate sample angle $\theta=77^{\circ}$, where $\theta=0^{\circ}$ puts the beam parallel to the sample surface. {\QCO} then occurs at  $2\theta=135^{\circ}$ $\theta=29^{\circ}$. The \gls{SO} reflections with odd $L$ arise from hypothetical body-centered-stacked stripes. Peaks with even $L$ are forbidden in the hypothetical body-centered-picture and can be thought of arising from stacking faults  \cite{Lee1997LSNO, Freeman2004LSNO}. The even $L$ peaks get enhanced in samples with hole concentrations away from the commensurate value, 1/3, due to increasing stacking faults \cite{Lee2001LSNO}.

However, the domain motions corresponding to $L=0$ versus $L=1$ are expected to be rather similar. Considering the motion of a $(0.36, 0.36)$ 2D spin modulation within a single NiO layer. There are two such NiO layers per unit cell and each layer is (by the crystal symmetry) equally well coupled with its neighboring layers in the same cell and in another cell. The only class of defect motion within a layer that would appear in the ideal (0.36, 0.36, 1) reflection but not in the (0.36, 0.36, 0) reflection is when a motion in one layer is canceled by an opposite motion in the other direction in the other layers. This type of coordinated motion is not physically plausible, as sliding the density waves against each other would cost far more energy than sliding in the same direction. It is also worth emphasizing that stripe order in nickelates is a bulk phenomenon so that the motion of a perfectly ordered domain will involve displacing the domain wall and the nearby domains. In addition, we note that the $(1/2-\epsilon/2, 1/2-\epsilon/2, 1)$ reflection cannot be reached with Ni $L$-edge resonance as the required {\Q} is larger than twice the momentum carried by the x-ray photons.

\begin{figure}
\includegraphics{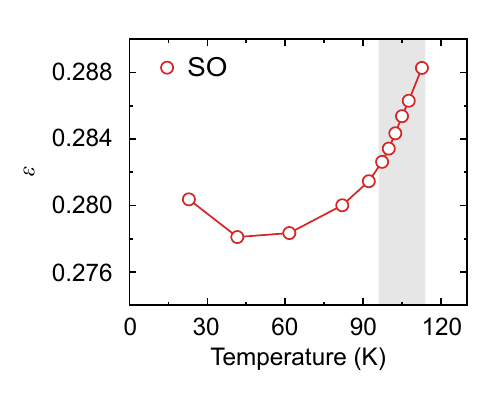}
\caption{Incommensurability derived from the \gls{SO} {\Q} vector as a function of temperature. The shaded areas indicate the onset temperature range for \gls{CO} and \gls{SO}.}
\end{figure}

For completeness we plot the incommensurability, $\epsilon$, derived from the \gls*{SO} in Fig.~S2, which is slightly larger than that from \gls*{CO}. The small difference in incommensurability may come from finite precision in determining the sample UB matrix and is comparable to our estimated uncertainty in alignment, noting that large angular motions and changing the x-ray energy and harmonics of the beamline undulator were required for the two measurements. We also present in Fig.~S3 the speckle patterns after different delay times at different temperatures for \gls{CO} and \gls{SO}.

\section{Calculations of cross-correlation function}

The cross-correlation function is used to quantitatively evaluate the speckle position changes. To so do, we first isolated the pure speckle patterns, which was done by dividing the raw image by the peak envelope that was estimated by fitting. Then the cross-correlation function is calculated through a pixel-to-pixel intensity correlation approach \cite{Chen2019LBCO}:

\begin{equation}
    A_{m,n} \ast B_{m,n} = \sum\limits_{m' = -M}^{M}\sum\limits_{n' = -N}^{N} A_{m',n'} B_{m+m', n+n'}.
    \label{eq:cross-correlation}
\end{equation}

\noindent where speckle patterns before and after temperature cycling are represented as two $M \times N$ matrices $A_{m,n}$ and $B_{m,n}$, respectively. Here, we chose an area of approximately $50 \times 50$ pixels as the input. The resulting cross-correlation matrix shows a peak at the center, the amplitude of which, $I(A, B)$, represents the similarity of these two images. Similarly, the auto-correlations can be obtained as $I(A, A)$ and $I(B, B)$. The normalized cross-correlation is then derived through:

\begin{equation}
    \xi_{\mathrm{CC}} = \frac{I(A, B)}{\sqrt{I(A, A) \times I(B, B)}}
    \label{eq:norm_cross-correlation}
\end{equation}

\begin{figure}
\includegraphics[width=1.\textwidth]{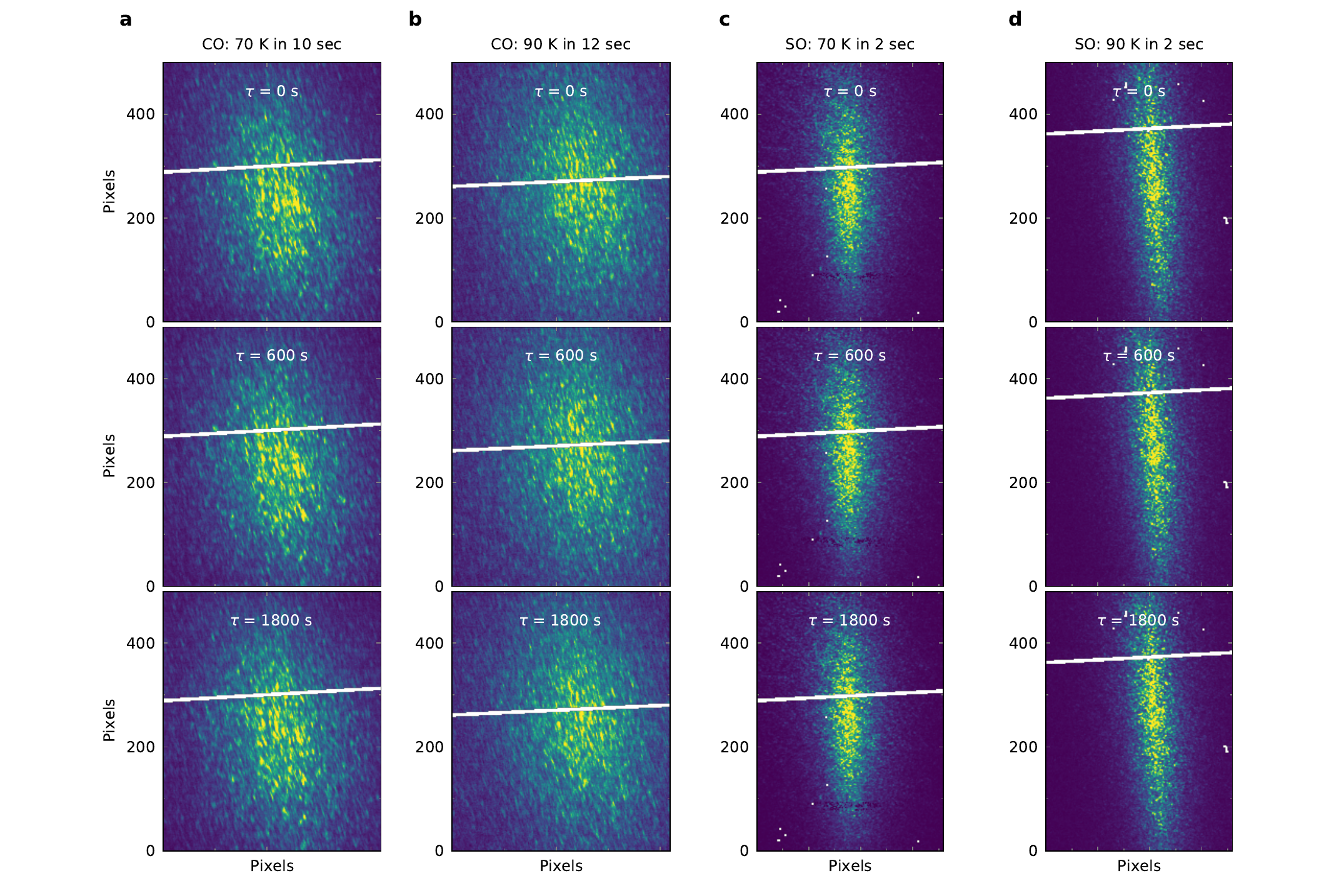}
\caption{Speckle patterns after different delay times at different temperatures for \gls{CO} and \gls{SO}.}
\end{figure}

\bibliography{refs}